\documentclass[a4paper,11pt]{article}
\pdfoutput=1 
\usepackage{jheppub, amsmath, amssymb,amsthm,dsfont,graphicx,mathtools,multirow,rotating,placeins,verbatim} 

\usepackage{tensor}
\usepackage[T1]{fontenc} 
\usepackage[all,color]{xy}
\usepackage{float}

\newcommand{\beq}{\begin{eqnarray}}
\newcommand{\eeq}{\end{eqnarray}}
\newcommand{\be}{\begin{equation}}
\newcommand{\ee}{\end{equation}}

\newcommand{\R}{\mathbb{R}}
\newcommand{\Av}{\mathcal{A}}

\newcommand{\D}{\mathcal{D}}
\newcommand{\F}{\mathcal{F}}
\newcommand{\Li}{\mathcal{L}}
\newcommand{\Ib}{\mathcal{I}}

\title{Renormalized electric and magnetic charges for $O(r^n)$ large gauge symmetries} 
\author[a]{Javier Peraza}

\affiliation[a] {\it   Facultad de Ciencias,  Universidad de la Rep\'ublica, \\ Igu\'a 4225, 
              esq. Mataojo, 11400 Montevideo, Uruguay.}
\emailAdd{jperaza@cmat.edu.uy}

\abstract{
In this work we present the construction of a renormalized symplectic form on an extended phases space where the higher order large gauge transformations (LGT) act canonically. The expressions of the sub$^n$-leading electric charges associated with each $O(r^n)$ LGT are then obtained, in agreement with the expressions previously proposed in \cite{Campiglia:2018dyi} by means of the tree-level sub$^n$-leading formulas. We also present the duality extension of the extended phase space, computing the full electromagnetic charge algebra, showing a tower of central extensions. 
}

\begin{document} 
\maketitle
\flushbottom

\section{Introduction}

Over the past years, the understanding of asymptotic symmetries in gravity and gauge theories has been deepened due to several results that relate them to soft theorems in field theory. The seminal works of Strominger and collaborators (e.g., \cite{Strominger:2013lka,He:2017fsb,He:2014laa,Cachazo:2014fwa,Lysov:2014csa,He:2014cra,He:2015zea,Kapec:2015ena}) showed that the well known Weinberg's soft theorem \cite{Weinberg:1965nx} can be understood as a Ward identity associated to an infinite dimensional symmetry group. The group is constructed via large gauge transformations (LGT) at null infinity. It implies infinite conservation laws in scattering processes from the past to the future asymptotic regions. 

In the case of Quantum Electrodynamics (QED), it was shown in \cite{Hamada:2018vrw} and \cite{Li:2018gnc} that for tree-level amplitudes, there exist an infinite number of soft theorems, each of them implying a conservation law for the tree level scattering process. Weinberg's soft photon theorem corresponds to the first level in the hierarchy, while Low's subleading soft photon theorem \cite{Low:1954kd, Low:1958sn} corresponds to the second level. 

The conserved quantities found in \citep{Strominger:2013lka} for the S-matrix constitute thus the first level in an infinite hierarchy of soft theorems. Seraj made a first approach towards higher orders in \cite{Seraj:2016jxi}, where an infinite number of conserved quantities are shown at spatial infinity, proportional to the multipole moments, and generated by specific large gauge transformations of order $O(r^n)$. At null infinity, Campiglia and Laddha showed in \cite{Campiglia:2018dyi} that (for tree-level scattering and restricting the radiative data space to a suitable subset) exists an infinite tower of conservation laws, such that at each level there is an infinite dimensional family of conserved charges, $Q^n_\epsilon$, labeled by functions on the sphere. The authors also presented evidence that the Ward identities associated with the level $n$ of the charges are equivalent to sub-$n$ soft photon theorems, along with the conservation laws within the classical theory. The non-abelian case is substantially harder since the charges up to level $n$ of the hierarchy do not form a closed algebra, as in the abelian case. In \citep{Campiglia:2021oqz}, a first step towards a classical derivation of the charge hierarchy in the non-abelian case is suggested. Some recent developments in celestial holography using Operator Product Expansion (OPE) tools \cite{Pasterski:2021rjz, Guevara:2021tvr, Guevara:2021abz} seem to be promising avenues in the study of asymptotic symmetries and the role of CCFT in flat holography for Yang-Mills and gravity. 

Working in terms of retarded coordinates $(u,r,x^1,x^2)$, the massless fields at the asymptotic region are determined by the limit $t:= r+u \rightarrow +\infty$ at constant $u$, where $t$ is the usual Minkowski time. This limit moves the Cauchy slices to a well-defined manifold, called the future null infinity and denoted by $\Ib^+$. \footnote{This convergence is point-wise equivalent to the limit $r\rightarrow +\infty$ at $u= cnt$, but taking $t\rightarrow +\infty$ is more natural since we are defining the charges in terms of Cauchy slices.} The topology of $\Ib^+$ is that of $\R \times S^2$ and its boundaries at $u = \pm \infty$ are denoted by $\Ib^+_{\pm}$ (they are diffeomorphic to $S^2$). 

The $r$-expansion of the LGTs at the bulk establishes a hierarchy of charges at the asymptotic region. $O(1)$ LGTs correspond to leading charges (for instance, by imposing a constant LGT we obtain the total electric charge of the system, \citep{Strominger:2017zoo}), while $O(r)$ LGT corresponds to sub-leading charges, see \citep{Lysov:2014csa,Campiglia:2016hvg}.  

The canonical derivation of conserved quantities at null infinity in the context of the classical theory at the leading and subleading imposes the following question: can the infinite tower of charges, associated with sub${}^n$-leading soft theorems, be canonically derived within the classical theory? One of the main problems that arise when studying $O(r)$ LGT is the divergent formulas for the charges when calculated from the usual phase space structure, both at null (e.g., \cite{Campiglia:2016hvg}) and spatial (e.g., \cite{Seraj:2016jxi}) infinities. In particular, the expressions for the symplectic form evaluated on an LGT at level $n$ (and therefore the charges) diverge in the $t \rightarrow +\infty$ and $u \rightarrow -\infty$ limits.

In this paper we provide a renormalization procedure that removes both divergences. Following ideas from \cite{Freidel:2019ohg}, we show that suitable boundary and corner terms exist for the symplectic form that renormalize the divergences while not changing the dynamics of the fields. This renormalization is minimal, in the sense that it cancels all the divergent terms while keeping \textit{unchanged} the finite ones. We define a subset of the radiative space and an extended phase space that contains all LGTs up to arbitrary order. This extended space has a symplectic structure, allowing us to calculate the electric-type charges. Finally, allowing the duality symmetry to act and extending the phase space with extra boundary gauge fields (e.g., \cite{Hosseinzadeh:2018dkh,Freidel:2018fsk,Geiller:2021gdk}), the magnetic analog of the electric hierarchy is also presented, as well as the full electromagnetic charge algebra.

The paper is organized as follows. In section 2 we review the asymptotic structure of Maxwell theory at null infinity. For simplicity, the charged matter consist of a massless complex scalar field coupled to the $U(1)$ gauge field. We also review the structure of the LGT for arbitrary order. In section 3, we revisit the derivation of asymptotic charges associated with leading and sub-leading soft photon theorems, defining an extended phase space and calculating the leading and sub-leading charges. Our derivation is along the lines of \cite{Campiglia:2018dyi}, but we place special emphasis on the symplectic structure, which will be used later. Section 4 contains the main result: we can renormalize the symplectic potential to have a finite value for every $O(r^n)$ LGT. Section 5 contains the derivation of magnetic charges, and the algebra of electromagnetic charges is presented. Finally, in section 6, we discuss the results and possible future directions.

\section{Preliminaries} \label{prelim}

This section reviews previous results on the asymptotic expansion of Maxwell fields at null infinity. 

\subsection{Radiative phase space}

Consider retarded coordinates $(u,r,x^a)$, in terms of which the Minkowski metric is
\be  \label{Mink}
ds^2 = -du^2 - 2dudr + r^2 q_{ab}dx^a dx^b.
\ee 
Indices $a,b,c,...$ will indicate sphere coordinates, while Greek indices $\mu,\nu,\sigma,...$ will indicate spacetime coordinates. The metric $q_{ab}$ is the standard round metric with constant curvature in the sphere $S^2$, with connection $D$. The limit $ r+ u =: t\rightarrow +\infty$ at fixed $u$ defines $\Ib^+$, ``scri plus'', a null hypersurface with the topology of $\R \times S^2$. Its boundaries are defined by the limits $u \to \pm \infty$, denoted by $\Ib^+_{\pm}$ respectively, and have the topology of a sphere.

We consider a massless charged scalar field $\phi$ coupled to the Maxwell field $\Av_\mu$ in Minkowski spacetime, with lagrangian
\be \label{Lagrangian}
\mathcal{L} = -\frac{1}{4} F_{\mu \nu} F^{\mu \nu} + \D_\mu \phi \overline{\D^\mu \phi},
\ee

and satisfying the field equations,
\beq
\nabla^\nu F_{\mu \nu} &=&  j_\mu, \\
\D_\mu \D^\nu \phi &=& 0,
\eeq
where $j_\mu = i e \phi \overline{\D_\mu \phi}   + c.c.$, with $\D_\mu \phi := \partial_\mu \phi - ie \Av_\mu \phi $, the gauge covariant derivative and $\nabla$ the metric covariant derivative. In retarded coordinates, Maxwell equations are 
\beq 
r^2 j_r &=& -\partial_r (r^2 F_{ru}) + D^a F_{ra}, \label{Maxr} \\
r^2 j_u &=& -\partial_r (r^2 F_{ru}) + \partial_u (r^2 F_{ru}) + D^a F_{ua},\label{Maxu} \\
j_a &=& \partial_r (F_{ua} - F_{ra})+ \partial_u F_{ra} + \frac{1}{r^2} D^b F_{ab}. \label{Maxa}
\eeq

Bianchi identities, $0=\partial_{[a} F_{bc]}$, are the integrability conditions for the electromagnetic strength tensor: there exists a one-form $\Av_\mu$ such that $F_{\mu \nu} = \partial_{[\mu} \Av_{\nu]}$. We will work in the harmonic gauge, $\nabla^\mu \Av_\mu = 0$ \footnote{We leave the study of the renormalization procedure in other gauges for future work. In particular, the light-cone gauge in the self-dual sector of Yang-Mills theory seems a promising avenue to extend the present results to non-abelian theories, \cite{Nagy:2022xxs}}, which in this particular coordinates implies

\be 
r^2 \partial_u  \Av_u + \partial_r (r^2 \Av_r) + r^2 D^a \Av_a = 0 .
\ee

We are interested in the symplectic structure and charges at $\Ib$, so we will need to take the $t \to +\infty $ limit, fixed $u$. Therefore, we need the $1/r$-expansion of the fields. The usual fall-offs for the electromagnetic tensor are (see \cite{Strominger:2017zoo} and \cite{Campiglia:2021oqz}):
\be 
F_{ru} = \frac{1}{r^2} F^{(-2)}_{ru} + o(r^{-2}), \quad F_{ar} = o(r^{-1}), \quad F_{au} = O(1), \quad F_{ab} = F^{(0)}_{ab} + o(1)
\ee
where it is understood that all the coefficients in the expansions are functions of $u$ and $x^a$. The fall-off for the scalar field is,

\be \label{decayphi}
\phi = \frac{\phi^{(-1)}}{r} + o(r^{-1}).
\ee
These expressions imply the following fall off's on the charge current:

\be
j_u = \frac{j_u^{(-2)}}{r^2} + o(r^{-2}) ,\quad j_a = \frac{j_A^{(-2)}}{r^2} + o(r^{-2}) ,\quad j_r = \frac{j_r^{(-2)}}{r^4} + o(r^{-4}).
\ee

Fall off's for $\Av_\mu$ compatible with the expansion above and the harmonic gauge condition are:

\be \label{decays}
\Av_a = \Av_a^{(0)} + o(1), \quad \Av_u = \Av_u^{(-1,\ln)} \frac{\ln r}{r} + o(r^{-1}), \quad \Av_r = o(r^{-1}).
\ee
The previous asymptotic behaviors are consistent with the field equations and the harmonic gauge condition. 

Using Maxwell equations, the scalar field equation, Bianchi identities and the harmonic gauge condition, we can solve all the components of the electromagnetic tensor and the scalar field in terms of $\Av_a^{(0)}$ and  $\phi^{(-1)}$ (see appendix A of \cite{Campiglia:2021oqz} for Yang-Mills case). These functions are the free data for the gauge field and the scalar field, respectively. To simplify notation, we will refer $\Av_a^{(0)}$ as $A_a$ and $\phi^{(-1)}$ as $\varphi$, respectively.

The hypothesis of ``tree-level'' decays for $A_a$ in the limits $u \rightarrow \pm \infty$,
\be \label{falloffA_a}
\partial_u A_a(u,x^1,x^2) = O(1/|u|^\infty),
\ee
that is, its decay is faster than that of any power $1/|u|^n$, implies the following fall-offs for the radiative data of a generic solution of Maxwell's equations \cite{Ashtekar:1981bq}, \cite{Campiglia:2018dyi},
\be \label{falloffFru}
F^{(-2)}_{ru}(u,x^1 ,x^2)= F^{-2,0}_{ru}(x^1 ,x^2) + O(1/|u|^{\infty}).
\ee 

For the massless charged scalar field, we assume a consistent fall off with \eqref{falloffFru} and the equations of motion, (\eqref{Maxr}, \eqref{Maxu} and \eqref{Maxa}) 
\be \label{falloffphi}
\varphi(u,x^1,x^2) = O(1/|u|^\infty).
\ee
This condition is of a technical nature, and it is imposed only for the sake of consistency of the equations. Our radiative phase space is thus defined in terms of the functions $A_a$ and $\varphi$,
\be \label{phasesp0}
\F_0 = \{ \left( A_A(u,x^a), \varphi(u,x^a) \right) : \partial_u A_a(u,x^1,x^2) , \varphi(u,x^1,x^2) = O(1/|u|^\infty) \}.
\ee

\subsection{$u$-expansions for fields}

From Maxwell equations and Bianchi identities, we can obtain recursive formulas for the coefficients in both $F_{ru}$ and $\epsilon^{ab}F_{ab}$ expansions in $r$ and $u$, where $\epsilon^{ab}$ is the area form of the sphere. By Bianchi identity $\partial_{[a}F_{ru]} = 0$, contracting with $D^a$ and the first two Maxwell equations, we have
\be
\Delta F_{ru} + \partial_r(\partial_r (r^2 F_{ru}) - 2 r^2 \partial_u F_{ru}) = r^2 \partial_u j_r - \partial_r (r^2 j_u) \label{Frurecursive}
\ee
where $\Delta$ denotes the Laplacian operator on the sphere, with metric $q_{ab}$. We assume that $F_{ru}$ can be expanded in an $r$-series, $F_{ru} = \frac{1}{r^2} \sum_{k=0}^{\infty} \frac{F_{ru}^{(-2-k)}}{r^{k}}$. By direct substitution in \eqref{Frurecursive},
\be \label{Maxorder}
2(k+1)\partial_u F_{ru}^{(-2-k-1)} + \left( \Delta + k(k+1) \right) F_{ru}^{(-2-k)}   = \partial_u j_r^{(-2-k)} + k j_r^{(-2-k)}.
\ee
From the assumed fall off \eqref{falloffFru}, and equation \eqref{Maxorder}, it is clear that the behaviour of $F_{ru}^{(-2-n)}$ in the limit $u\to -\infty$ is
\be \label{frudecay}
F^{(-2-n)}_{ru} = \sum_{j=0}^n u^j F_{ru}^{(-2-n,j)}(x^a) + r_n(u,x^a),
\ee
where each of the $F_{ru}^{(-2-n,0)}(x^a)$ is a function on the sphere, and $r_n$ some function with an $O(1 / u^{\infty})$ decay (analogous expansion can be done in the limit $u \rightarrow + \infty$). We can solve order by order recursively in terms of the current and these free functions. As a reference, the full expression for $F_{ru}$ is
\be
r^2 F_{ru} = \sum_{k \geq 0} \frac{1}{r^k} \sum_{j=0}^k u^j F_{ru}^{(-2-k,j)}(x^a)  + ...
\ee

The same analysis can be carried out for the function $\epsilon^{ab} F_{ab}$, obtaining the following equation,

\be 
2 \epsilon^{ab} D_a j_b = 2\partial_u \partial_r \epsilon^{ab} F_{ab} - \partial_r \partial_r \epsilon^{ab} F_{ab} - \frac{1}{r^2} (\Delta \epsilon^{ab} F_{ab} + 4\epsilon^{ab} F_{ab}), \label{FABrecursive}
\ee
and by performing the $r-$ and $u-$expansions we obtain a recursive formula for the coefficients in $\epsilon^{ab} F_{ab}$ (see \autoref{ap_rec} for details). In section \eqref{magnetictower}, we will use these results.

\subsection{Variation space}

We now turn to the large gauge transformations (LGT) on the variation space. The usual formulas for the gauge symmetries,

\be \label{finiteaction}
\Av_\mu \mapsto \Av_\mu + \partial_\mu \epsilon ,\quad \phi \mapsto e^{-ie \epsilon} \phi
\ee
establish the following action for variations of the fields,
\be \label{varact}
\delta_\epsilon \Av_\mu = \partial_\mu \epsilon , \quad \delta_\epsilon \phi = -ie \epsilon \phi.
\ee

The variations allowed in our radiative phase space are tangent to $\F_0$, i.e., that maintain the fall-offs of the fields. By the definition of finite symmetry, given a gauge symmetry generator $\epsilon$ we see that $\partial_\mu \epsilon$ must have the same fall offs as $\Av_\mu$:

\be \label{decaysepsilon}
\partial_a \epsilon = O(1), \quad \partial_u \epsilon = o(1), \quad \partial_r \epsilon = o(r^{-1}) 
\ee

We study the global symmetries as arising from the residual LGT, and by the choice of harmonic gauge, are solutions to the wave equation,

\be 
\square \epsilon = 0.
\ee
This equation can be solved up to order $O(r^{-1})$ (see Appendix A in \cite{Campiglia:2016hvg}),

\be \label{epsilon0}
\epsilon(u,r,x^1,x^2) = \epsilon_0(x^1 ,x^2) + O(\ln (r) /r).
\ee

\subsection{Higher order LGT}

We are interested in relating the LGTs containing higher orders in $r$ with the charges that arise from sub$^n-$leading soft photons theorems. The usual mode expansion reasoning in the soft theorem derivation suggests that for a sub$^n-$leading soft photon, we need to look for an LGT $\Lambda$ whose $O(1)$ in the $r-$expansion behaves as $u^n$. This asymptotic behavior of the gauge generator must be compatible with the harmonic gauge and, therefore, implies an $O(r^n)$ leading behavior, as we show below by solving $\square \Lambda = 0$. 

Consider the following $r-$expansion for a $O(r^n)$ large gauge parameter,
\be
\Lambda(u,x^a) = r^n \epsilon^{(n)} + \sum_{k = 0}^{n-1} r^k \epsilon^{(k)} + \frac{\ln r }{r} \epsilon^{ln} + O(r^{-1}),
\ee
where $\epsilon^{(i)} = \epsilon^{(i)}(u,x^1,x^2)$. We have $\square \Lambda = 0$, which in retarded coordinates reads,
\beq
0 &=& -6 r^{n-1} \partial_u \epsilon^{(n)}  + \sum_{k = - 1}^{n-2} r^k \left( \Delta \epsilon^{(k+2)} - 2 (k+2) \partial_u \epsilon^{(k+1)} + (k+2)(k+3) \epsilon^{(k+2)} \right) \nonumber \\
&& + \frac{\ln r }{r^3} \Delta \epsilon^{(ln)} + \frac{2}{r^2} ( \Delta  \epsilon^{(0)} - \partial_u \epsilon^{(ln)} ) + \frac{1}{r^3} \epsilon^{(ln)} + ....  \label{harmonicgauge}
\eeq
The first term in \eqref{harmonicgauge} implies that $\epsilon^{(n)}$ is a free function on the sphere. Next, we have a recursive equation on the successive coefficients:
\be \label{lgtrec}
2 (k+1) \partial_u \epsilon^{(k)} = \Delta \epsilon^{(k+1)} + (k+1)(k+2) \epsilon^{(k+1)}
\ee

Integrating \eqref{lgtrec} and fixing each integration constant to zero in each step gives an LGT of order $O(r^n)$ generated by $\epsilon \equiv \epsilon^{(n)}$, which we will call $\Lambda^n_\epsilon$. If the integration constants are non-zero, each one of them will be a free $S^2$ function that contributes linearly with an LGT of corresponding order:
\be \label{lgtlin}
\Lambda_{\alpha} =  \Lambda^n_{\epsilon_n} + \Lambda^{n-1}_{\epsilon_{(n-1)}} + ...,
\ee
where $\alpha = \{\epsilon_j\}_j$ is the sequence of integration constants $\epsilon_{j}$ in the equation \eqref{lgtrec}, that are free $S^2-$functions, each one generating an $O(r^j)$ $LGT$. We will call an LGT ``pure'' if only one free function generates it. When using the notation $\Lambda^m_{f}$, subscripts indicate the generating function or sequence of functions, and superscripts indicate the leading term in the $r-$expansion, if the generating function is not a sequence.

Some remarks are in order. First, one implication of equation \eqref{lgtrec} for a pure $O(r^n)$ LGT is the following property:
\be
\epsilon^{(n-1)} = O(u),\quad ..., \quad \epsilon^{(k)} = O(u^{n-k})
\ee

This shows that the order $O(r^n)$ is necessary for a $u^n$ asymptotic behavior at order $O(r^0)$ for the LGT, as was stated at the beginning of the section. Second, the term $\ln(r ) /r$ is needed for the $O(r^0)$ to be consistent; otherwise, we would get $\Delta \epsilon^{(0)}=0$, and since we are in a sphere, that would give a trivial function. Third, the non-trivial fact that equation \eqref{lgtrec} resembles the form of equation \eqref{Maxorder}, but it presents crucial differences in the constants multiplying the functions. This similarity between the recursive expressions is useful when showing the Ward identity equivalence with the sub$^n-$ soft theorems.

\section{Leading and Subleading charges}

In this section, we review the phase space construction and the symplectic charges in the case where the LGTs are $O(r)$. We leave the renormalization procedure for the next section, focusing exclusively on the first step of the phase space extension and the recovery of the charges.

\subsection{Extended phase space}

The usual phase space, \eqref{phasesp0}, contains the physical information regarding the leading order charges, restricted to $O(r^0)$ LGT. Their usual expressions are (\cite{Strominger:2013lka},\cite{Strominger:2017zoo}):
\be 
Q_{\epsilon_0} = \int_{S^2} \sqrt{q}  \epsilon_0 \int_{\R} \partial_u F^{(-2)}_{ru} du ,
\ee
where $\epsilon_0$ is a function on the sphere. The fall-offs \eqref{decays} are not preserved by an $O(r^1)$ LGT as soon as we allow higher order LGT (through its action \eqref{finiteaction}) and therefore, the variations are no longer tangent to the radiative phase space $\F_0$. We expand the phase space in these extra directions by first extending the vector potential sector in \eqref{phasesp0}. Consider the following space:
\be 
\F_1 = \F_0 \times \{ \psi(x^1,x^2) : \psi \in C^{\infty}(S^2)\}
\ee 

We define the new vector potential as $\hat{\Av}_\mu = \Av_\mu + \partial_\mu \Lambda^1_\psi$, where $\Av_\mu$ is the vector potential that has $A_a$ as initial data (from section \eqref{prelim}) and $\Lambda^1_\psi$ is the pure $O(r)$ LGT generated by $\psi$. Observe that this definition is indeed consistent, since $\partial_{[\mu} \partial_{\nu ]}\Lambda^1_\psi = 0$ and thus $a$ makes no contribution to the electromagnetic tensor, i.e. $\hat{F} = F$. \footnote{This feature in the abelian case is in sharp contrast to the non-abelian case, where the linear extension was studied in \cite{Campiglia:2021oqz}} Observe also that the harmonic gauge condition is trivially satisfied for the extended electromagnetic potential \footnote{It is left for future works to study the phase space extension in more general gauges, and whether it changes the structure.}.

Given a general $O(r)$ LGT, $\Lambda_{\{ \epsilon_1 , \epsilon_0\} }$, the variations generated by it on $\F_1$ are split in terms of the $S^2$ free functions $\epsilon_1$ and $\epsilon_0$ corresponding to order $O(r)$ and order $O(1)$ in the $r-$expansion respectively (see \eqref{lgtlin}):
\be 
\Lambda_{\{ \epsilon_1 , \epsilon_0\}}  = r \epsilon_1 + (\epsilon_0 + u\frac{1}{2}(\Delta + 2)\epsilon_1) + O(\frac{\ln r}{r})
\ee 

The action on the phase space $\F_1$ comes from the identity $\delta_{\Lambda_{\{ \epsilon_1 , \epsilon_0\} }} \hat{\Av}_\mu = \partial_\mu \Lambda_{\{ \epsilon_1 , \epsilon_0\} }$, which after the splitting reads:
\beq
\delta_{\Lambda_{\{ \epsilon_1 , \epsilon_0\} }} A_a &=& \partial_a \epsilon_0, \\
\delta_{\Lambda_{\{ \epsilon_1 , \epsilon_0\} }} \psi &=& \epsilon_1.
\eeq

Allowing a $O(r)$ LGT in the massless field sector also implies a change in the massless field $\phi$. The equations of motion are invariant under the simultaneous change
\be 
\Av_\mu \mapsto \hat{\Av} = \Av_\mu + \partial_\mu \Lambda^1_\psi , \quad \phi \mapsto \hat{\phi} = e^{-ie \Lambda^1_\psi}\phi.
\ee

Since the finite gauge symmetry involves a product $e^{-ie \Lambda^1_\psi} \phi$, we can define an extended field $\hat{\phi} = e^{-ie \Lambda^1_\psi} \phi$, where $\psi$ is the free $S^2$ function now generating a phase for the scalar field, while $\phi$ is the massless field with the usual fall off, with $\varphi \in \F_0$ as free data. The covariant gauge derivative is given by
\be
\hat{\D}_\mu := \partial_\mu - ie\hat{\Av}_\mu,
\ee
from where we have that the new current $\hat{j}_\mu$ maintain its original form,
\be 
\hat{j}_\mu = ie \hat{\phi} (\hat{\D}_\mu \hat{\phi})^* + c.c. = ie \phi (\D_\mu \phi)^* + c.c.,
\ee
The consistency of the action of the $O(r)$ LGT action on $\hat{\phi}$ with the splitting of the extended phase space implies
\be 
\delta_{\Lambda_{\{ \epsilon_1 , \epsilon_0\} }} \varphi = - ie\epsilon_0 \varphi.
\ee

This type of extension of the phase space and the dressing of the fields is part of a more general procedure, using ``Goldstone modes'' on the boundary, that has been introduced both in the context of gauge theories and gravity (see \cite{Donnelly:2016auv,Freidel:2018fsk,Ciambelli:2021nmv, Campiglia:2021oqz} and references therein).

\subsection{Calculation of leading and subleading charges}

This subsection reviews the covariant phase space procedure for calculating charges associated with a gauge transformation generated by $\epsilon$. Consider the Lagrangian \eqref{Lagrangian}, in our extended phase space we have the usual symplectic potential current,

\be \label{asymppot}
\theta^\mu (\delta) = \sqrt{g} \left( \hat{F}^{\mu \nu} \delta \hat{\Av}_\nu + \hat{\D}^\mu \hat{\phi} \delta \bar{\hat{\phi}} + c.c. \right),
\ee
and the symplectic current by taking the exterior derivative in the phase space,
\be 
\omega^\mu (\delta , \delta') = \delta \theta^\mu (\delta') - \delta' \theta^\mu (\delta) - \theta ([\delta , \delta']).
\ee
The symplectic form is obtained by integrating the symplectic current over $\Sigma_t$, $\Omega (\delta, \delta') = \int_{\Sigma_t} \omega^{\mu} (\delta, \delta')  d S_\mu$. We evaluate it on a variation generated by a general LGT ($\Lambda_{\epsilon_1 , \epsilon_0}$) and an admissible variation (denoted by $\delta$), obtaining an expression for the charge, 
\be \label{defcharge1}
\delta Q_{\Lambda_{\epsilon_1 , \epsilon_0}} = \Omega(\delta , \delta_{\Lambda_{\epsilon_1 , \epsilon_0}}) = \int_{\Sigma_t} \omega^\mu (\delta, \delta_{\Lambda_{\epsilon_1 , \epsilon_0}}) dS_\mu
\ee
where the integrals are taken over a $t= cnt$ surface. As it was shown in \cite{Campiglia:2016hvg}, one could find the leading and subleading charges (consistent with the Ward identities) by taking the limit $t = r + u \rightarrow +\infty$ at constant $u$,
\be \label{chargefull}
Q_{\Lambda_{\epsilon_1 , \epsilon_0}} = \lim_{t \rightarrow \infty} \int_{\Sigma_t} (\partial_r - \partial_u ) ( r^2 \Lambda_{\epsilon_1 , \epsilon_0} \hat{F}_{ru} ) dx^2 du,
\ee
and considering the finite part in the limit. By counting orders in $t$, it is straightforward to see that the expression \eqref{chargefull} contains divergent terms; therefore, the definition of the charge at the limit is ill-defined. In what follows, we drop the hat $\hat{•}$ in $F_{ru}$ since it is the same field as in the radiative space.

As we previously mentioned, the main result of this paper is that we can define a procedure to renormalize the symplectic potential and get rid of the divergent terms in \eqref{chargefull} for any arbitrary higher order $O(r^n)$. This will be the content of the next section, while in the remainder of this section, we motivate the renormalization in the particular case of the extension for $n=1$.

Since we can trace back the divergences to the symplectic potential, due to varying with $\delta_{\Lambda_{\epsilon_1 , \epsilon_0}}$, our starting point is to compute the symplectic potential on the hypersurfaces $\Sigma_t$,
\be \label{thetaextended1}
\theta^t (\delta) = \underbrace{\sqrt{q} \left(r^2 F_{ru} (\delta A_r - \delta A_u) + q^{bc} F_{ub} \delta A_c \right)}_{\theta_0^t} + \underbrace{\sqrt{q} (\partial_r - \partial_u)(r^2 F_{ru} \delta \Lambda^1_\psi)}_{\theta_1^t}, 
\ee
where we did not write the total derivative $r^2 D_c( \sqrt{q} q^{bc} F_{ub} \delta \Lambda_\psi^1)$, since it vanishes after integration on $\Sigma_t$. The first term can be regarded as the radiative phase space symplectic potential, $\theta_0^t$, while the second term is the new extended term, which we will call $\theta_1^t$. 

The term $\theta_0^t(\delta)$ will contribute to the symplectic form (by integrating by parts and using the equations of motion) as usual,
\be
\omega_0^t(\delta,\delta') = \sqrt{q} q^{bc}\delta  F_{ub} \wedge \delta' \Av_c + \sqrt{q} r^2 \delta F_{ru} \wedge \delta' ( \Av_r  -  \Av_u  ),
\ee

The term $\theta_1^t(\delta)$ presents the divergence: the action of $\partial_u$ on $\delta \Lambda_\psi^1$ leaves an $O(r)$ term, which in turns imply a $t$ factor when changing variables from $(u,r,x^1 ,x^2)$ to $(t,r,x^1, x^2)$. In the next section we show a systematic approach for renormalizing such terms while keeping the finite ones unchanged (i.e., the minimal subtraction to make the expressions finite). For now, we assume that we can drop the divergent term and that the expression we get also has a finite limit $u \to -\infty$. Assuming the above, we find the following expression for the renormalization of $\theta_1^t(\delta)$,
\be
\theta_1^{ren,t} (\delta) = \sqrt{q} \left( D^a j_a^{(0)} - \frac{u}{2} \Delta \partial_u F_{ru}^{(-2)} \right) \delta \psi ,
\ee
where $ren$ stands for ``renormalized''. The symplectic current is split then,
\be 
\omega^{ren,t}(\delta,\delta') = \omega_0^t(\delta,\delta') + \omega_1^{ren,t} (\delta,\delta'),
\ee
where the last term comes from the exterior derivative of $\theta_1^{ren,t} (\delta)$, and the total symplectic form on $\Ib^+$ is well defined (by taking $t \rightarrow + \infty$) ,
\be
\Omega^{ren}(\delta , \delta') =  \int_\Ib \omega^{ren}(\delta,\delta') = \int_\Ib \omega_0(\delta,\delta') + \int_{S^2}\sqrt{q} (\delta F_{ru}^{(-3,0)} \wedge \delta' \psi) .
\ee
The last term comes from the value of $F_{ru}^{(-3,0)}$ in \eqref{frudecay}, which can be seen as the value of the following limit (see \cite{Campiglia:2018dyi} for details):
\be
F_{ru}^{(-3,0)} = \lim_{u\to -\infty} F_{ru}^{(-3)} - u F_{ru}^{(-3,1)} = \int_\R \left( D^a j_a^{(0)} - \frac{u}{2} \Delta \partial_u F_{ru}^{(-2)} \right)  du,
\ee
where the contribution from $u=+\infty$ in integral zero due to the absence of massive charges ($F^{(2)}_{ru} (u = +\infty ,x^1 , x^2) = 0$). Since $\partial_u F_{ru}^{(-2)}$ decays faster than any polynomial in $u$, the above integral is convergent. Observe that $F_{ru}^{(-3,0)}$ is the canonical conjugate to $\psi$.

Next, we compute the leading and subleading charges. Taking $\delta'$ to be a large gauge transformation, and $\delta$ any arbitrary admissible variation (compatible with $\F_1$), we calculate the charge associated to any LGT $\Lambda_{\{ \epsilon_1 , \epsilon_0\} }$ by equation \eqref{defcharge1}. Since $F_{\mu \nu}$ is invariant under $\delta_{\Lambda_{\{ \epsilon_1 , \epsilon_0\} }}$ and $\Lambda_{\{ \epsilon_1 , \epsilon_0\} }$ is not affected by $\delta$ \footnote{This again is in contrast with the non-abelian case, where the harmonic gauge condition implies a field dependent LGT's.}, the calculation is straightforward,
\be
Q_{\Lambda_{\{ \epsilon_1 , \epsilon_0\} }} = \int_{S^2} \sqrt{q} \left( \epsilon_0 F_{ru}^{(-2,0)} + \epsilon_1 F_{ru}^{(-3,0)} \right)dx^2 =: Q^0_{\epsilon_0} + Q^1_{\epsilon_1} ,
\ee
where we also used \eqref{frudecay} in the radiative space sector, and $Q^i_{\epsilon_i}$, with $i=0,1$, denotes the leading and subleading charges, respectively. 

The charge $Q^0_{\epsilon_0}$ is the usual for a $O(1)$ large gauge symmetry, while the second term is the one obtained in \cite{Campiglia:2016hvg} and \cite{Campiglia:2018dyi}. In both cases, we obtained ``corner'' charges. They depend on the values of the fields at the boundary of $\Ib$, which is by itself the boundary of the domain we started with (as in \cite{Donnelly:2016auv, Freidel:2018fsk}). 

\section{Tower of asymptotic charges}

In this section, we derive an infinite hierarchy of charges from a symplectic form in an extended phase space that contains enough degrees of freedom to allow for $O(r^n)$ LGTs for arbitrary $n$. Certain difficulties in defining the symplectic potential arise, in particular, the appearance of several divergent integrals, as shown in the previous section. The renormalization procedure we apply is based on \cite{Freidel:2019ohg}.
 
First, we define the extended phase space and show the type of divergences we have, both in the $t \rightarrow +\infty$ and $u \rightarrow + \infty$ limits inside the expression \eqref{chargefull}. Then, we proceed to prescribe a renormalization on the symplectic potential that will lead to the correct expression for the charges while the symplectic form remains finite.
 
\subsection{Extended phase space and charges} \label{extended phase space and charges}

Let $\mathcal{S}$ be the space of sequences $\{\psi_i \}_{i > 0}$ of functions $\psi_i : S^2 \to \R$ such that only finitely many are non-zero \footnote{In what follows we assume that the sequences of functions have this property unless stated otherwise}. Given a sequence $\Psi \in \mathcal{S}$, we define the LGT associated to the sequence as 

\be
\Lambda_{\Psi} := \sum_{i > 0} \Lambda^i_{\psi_i} 
\ee
where each $\Lambda^i_{\psi_i}$ is a pure $O(r^i)$ LGT associated to $\psi_i$, in the sense of section \eqref{prelim}. Observe that the sum is finite for every $\Psi \in \mathcal{S}$. We define the extended phase space as the following set,

\be
\F_{\infty} = \F_0 \times \mathcal{S},
\ee
with the extended electromagnetic potential and scalar field are defined as 

\be 
\hat{\Av}_\mu = \Av_\mu + \partial_\mu \Lambda_\Psi, \quad \hat{\phi} = e^{-ie \Lambda_\Psi} \phi,
\ee
where $\Av_\mu$ and $\phi$ are the vector potential and the scalar field generated by the free data $A_a$ and $\varphi$ from the space $\F_0$, respectively.

The admissible variations $\delta$ of this phase space are such that when acting on the degrees of freedom parametrized by $\Psi$, it satisfies $\delta \Psi \in \mathcal{S}$. This property is not restrictive regarding the variations, as shown below. 

Given a sequence $\varepsilon = \{\epsilon_0 , \epsilon_1 , ..., \epsilon_i, ...\}$ of free $S^2$ functions, such that $\{\epsilon_i \}_{i > 0} \in \mathcal{S}$, consider the LGT associated to it, $\Lambda_{\varepsilon} = \Lambda^0_{\epsilon_0} + \sum_{i > 0} \Lambda^i_{\epsilon_i} $. The variation generated by this LGT acts on $\F_{\infty}$ by acting in $\Av_\mu$ with its $O(r^0)$ free function and by acting on $\alpha$ on each sequence term,

\be \label{varaction}
\delta_{\Lambda_\varepsilon} A_A = \partial_A \epsilon_0, \quad \delta_{\Lambda_\varepsilon} \varphi = -ie\epsilon_0 \varphi , \quad \delta_{\Lambda_\varepsilon} \Psi = \{ \epsilon_i\}_{i>0}
\ee
This structure is the same as in the previous section, extended to contain any order in the $r-$expansion. 

We can write the full symplectic potential, equation \eqref{asymppot}, and proceed in the same way as in the previous section, obtaining the expression \eqref{thetaextended1}. In this case, $\Lambda_\Psi$ is in place of $\Lambda_\psi^1$ and the split of the symplectic potential in the radiative phase space contribution and the extended part is given by
\be
\theta^t_{\infty} (\delta) = \sqrt{q} (\partial_r - \partial_u)(r^2 F_{ru} \delta \Lambda_\Psi), 
\ee
where the $\infty$ stands for the extension to all orders in $r$.

Given $\delta$ and $\Lambda_\Psi$, let us calculate the symplectic potential evaluated at $\delta$. Consider the integral, 
\be
\Theta_{t,\infty}(\delta) = \int_{\Sigma_t} \sqrt{q} (\partial_r - \partial_u)(r^2 F_{ru} \delta \Lambda_\Psi) dx^2 du,
\ee
and observe that the term inside the integral is divergent in the limit $t\rightarrow +\infty$ with the same order as the highest power of $r$ in $\delta \Lambda_\alpha$. Our aim in this section is to understand better this integral. For brevity let us call,
\be \label{rhos}
\rho_{k}(\delta) =   \sum_{i=k}^{+ \infty} F_{ru}^{(-2 + k - i)} \delta \Lambda_\Psi^{(i)},
\ee
where $\delta \Lambda_\Psi^{(i)}$ is the coefficient corresponding to $r^i$ in the $r-$expansion of $\delta \Lambda_\Psi$. $\rho_k (\delta) $ is thus the $O(r^k)$ coefficient in the expansion of the term inside the brackets. Upon direct computation, we have, 
 \be
\Theta_{t,\infty}(\delta)  = \int_{\Sigma_t} \sqrt{q} \sum_{k=1}^{\infty} \left( k r^{k-1} \rho_k (\delta)  - r^k \partial_u \rho_k (\delta) \right) dx^2 du,
\ee
which after we substitute $r = t-u$, gives,
\be
\Theta_{t,\infty}(\delta) = \sum_{j=0}^\infty t^j \int_{\Sigma_t} \theta_j^t(\delta) dx^2 du,
\ee
for some $t-$independent functions $\theta_j^t(\delta)$. This gives us a $t-$expansion of the symplectic potential. In the next subsection, we show that these divergences can be renormalized by adding total variations and total derivatives (corner) terms to the symplectic potential.

Assuming such a procedure can be done, we are left with the $O(t^0)$ term, which satisfies the identity,
\beq
\Theta^\Ib_\infty (\delta) &:=&  \lim_{t \rightarrow +\infty} \Theta_{t,\infty}(\delta)  = \int_{\Ib} \sqrt{q} \sum_{k=1}^{\infty} \left( k (-u)^{k-1} \rho_k (\delta) - (-u)^k \partial_u \rho_k (\delta) \right) dx^2 du \nonumber \\
&=& - \int_{\Ib} \partial_u \left( \sqrt{q} \sum_{k=1}^{\infty} (-u)^{k} \rho_k (\delta)  \right) dx^2 du, \label{symp_pot_divergent}
\eeq
which gives us a boundary term. The charges associated with higher order LGT can be directly computed using the identity $\delta Q_{\Lambda_\varepsilon} = \Omega^\Ib_\infty (\delta , \delta_{\Lambda_\varepsilon})$,
\be
Q_\epsilon = \int_{\Ib} \partial_u \left( \sum_{k=1}^\infty (-u)^{k} \rho_k (\delta_{\Lambda_\varepsilon}) \right) du d^2x. \label{divergent_charge}
\ee

When evaluating the term in the brackets in the last line of \eqref{symp_pot_divergent} at $u=+\infty$, we use the hypothesis that $F_{ru}= 0$ at $\Ib^+_+$. When evaluating at $\Ib^+_-$, we run into divergences. Since the general behavior of $\rho_k (\delta)$ admitted by \eqref{falloffA_a}, \eqref{falloffphi} and \eqref{Frurecursive} \footnote{Remember that the fall off \eqref{falloffFru} is a consequence of these equations.} near spatial infinity is polynomial in $u$ plus a $O(1/|u|^{\infty})$ remainder, we have that the above expression for $\Theta_{t} (\delta)$ is not well defined. By the renormalization procedure of the next subsection, we will be able to regularize the above expression, keeping only the $O(u^0)$ in $\rho_0 (\delta)$,

\be
\Theta^\Ib_\infty (\delta)  =  \int_{S^2}  \sqrt{q} \sum_{i=1}^{\infty} F_{ru}^{(-2-i,0)} \delta \psi_i dx^2 du,
\ee
where $F_{ru}^{(-2-i,0)}$ are the $O(u^0)$ of $F_{ru}^{(-2-i)}$.

The renormalization procedure of the next subsection has to address the previous two divergences: the $t$ divergence from the limit to $\Ib$, and the $u$ divergences in the integrals over $\Ib$. We leave it as a future work to understand the physical meaning of the boundary and corner terms in the context of covariant phase space quantities.

We end this subsection with some remarks regarding previous works. The idea in \cite{Campiglia:2016hvg} is to relate the divergent terms to the conserved quantities, obtaining a "projected out" charge equal to the $t^0$ term, while the discarded terms are proportional to lower order charges. While this is the case for the $O(r)$ subleading charge (the $O(t^1)$ part in the charge is proportional to $Q_{\epsilon_0}$), there is, however, a remaining divergent term in the $O(r^2)$ that leads to an unresolved tension, in particular the $O(t^1)$ term is not proportional to any lower order charge. This tension is solved once we renormalize the symplectic potential. 
 
Regarding the concrete expressions of the charges, the order $O(u^0)$ of $Q_{\Lambda_\varepsilon}$ in equation \eqref{divergent_charge} is exactly what was presented in \cite{Campiglia:2018dyi}. This is equivalent to prove that the $O(u^0)$ coefficient of $Q_{\Lambda_\varepsilon}$ is $ \sum_{k = 1}^{+\infty} \int_{S^2} \epsilon_k F_{ru}^{-2-k,0} d^2x$ (we are considering $\epsilon_0 = 0$, only higher order charges).

As it stands, \eqref{divergent_charge} diverges, due to the orders of $u^n$ that are involved in the integral. If we want to write the charge as a corner integral on the sphere at $u \rightarrow -\infty$, we should inspect the $O(u^0)$ term, corresponding to the finite limit term.

Here, we take the $u-$decay in the remainder functions $r_i$ in equation \eqref{frudecay} as faster than any polynomial decay. Therefore, inspecting the expressions for $\Lambda_\varepsilon^{(k)}$ and $F_{ru}^{-2+k-i}$, we see that each $\rho_k (\delta_{\Lambda_\varepsilon})$ has at least order $u^0$, therefore the term in the sum contributes with at least $u^k$. The only term with a possible $u^0$ order is thus $\rho_0 (\delta_{\Lambda_\varepsilon})$,

\be 
\rho_0 (\delta_{\Lambda_\varepsilon}) =   \sum_{i=1}^{\infty} \Lambda_\varepsilon^{(k)} F_{ru}^{(-2+k)}.
\ee
Again, a close inspection in the $u-$ expansion of the functions shows that the order $u^0$ is given by the sum of the products $\epsilon_{k} F_{ru}^{(-2-k,0)}$.

\subsection{Regularization procedure}

In this subsection, following \cite{Freidel:2019ohg}, we will renormalize the symplectic potential for QED in the extended phase space to eliminate the divergences. The idea is to write the higher order terms in the $t$ component of the symplectic potential as boundary plus corner terms and to subtract them from the original expression, thus obtaining a finite result in the limit $t \rightarrow \infty$.

From the first variation of the Lagrangian \eqref{Lagrangian}, we have,

\be 
\delta \Li = E^{\mu} \delta \hat{\Av}_\mu + E \delta \hat{\phi} + \partial_\mu \theta^\mu (\delta),
\ee
where $E^\mu$ and $E$ are the field equations for $\hat{\Av}_\mu$ and the massless scalar, respectively. By taking the retarded coordinates ${u,t,x^1,x^2}$ on Minkowski spacetime, we write the previous equation on-shell and obtain an equation for $\partial_t \theta^t (\delta)$,

\be \label{radialeq}
\partial_t \theta^t (\delta) = \delta \Li - \partial_u \theta^u (\delta) - D_a \theta^a (\delta).
\ee

We will assume that all the functions have $t$ and $u$ expansions around $t = +\infty$ and $u = \pm \infty$, as is the case for $F_{ru}^{(2)}$, $A_a$ and $\varphi$ (cf. equations \eqref{falloffFru}, \eqref{decays} and \eqref{decayphi}).
%

Consider the derivation of the divergent part of the symplectic potential done in the previous section but now applied to our extended phase space, i.e., starting from,

\be 
\theta^\mu (\delta) = \sqrt{q} r^2\left( F^{\mu \nu} \delta \hat{\Av}_\nu) + \overline{\hat{\D}^\mu} \hat{\phi} \delta \hat{\phi} + c.c.\right),
\ee
the general form for the symplectic potential on Cauchy slices at constant $t$ is the following, \footnote{In the following equations we write the explicit dependence of the functions on variations and coordinates.} 
\be \label{generaltheta}
\theta^t (\delta) = Y_0(\delta)(u,t,x^a) +  \sum_{i=1}^{\infty} t^i Y_{i}(\delta)(u,x^a).
\ee
where $Y_0(\delta)(u,t,x^a)$ is such that $\lim_{t \rightarrow +\infty} Y_0(\delta)(u,t,x^a) = Y_0(\delta)(u,x^a)$ is a well defined function on $\Ib^+$. We introduce the renormalized symplectic potential as $\theta_{ren}^t := \theta^t - H_{ren}$, where $H_{ren}$ satisfies the following equation,

\be \label{rnsympot}
\partial_t \theta^t (\delta) - \partial_t H_{ren}(\delta) = K(\delta)(u,t,x^a),
\ee 
where $K$ is such that its limit when $t \rightarrow +\infty$ vanishes. In general, $K$ and $H_{ren}$ are not uniquely determined by the previous equation. The natural prescription for $H_{ren}$ to resolve the divergences is the following, 
\be 
H_{ren}(\delta) =  \sum_{i=1}^{+\infty} t^{i} Y_i(\delta)(u,x^a) + C(\delta)(u,x^a),
\ee 
where $C(u,x^a)$ is a function to be determined. Observe that $H_{ren}$ has the same order as $\theta^t$ in the $t-$expansion, and that the divergences in the $t$ parameter are canceled, so $\theta^t_{ren}$ converges in the limit $t \to +\infty$. The coefficients $Y_i$ are obtained from the integration of the terms in the variation of the lagrangian and the total derivative of the symplectic potential in equation \eqref{radialeq}, on $\{t= cnt\}$ surfaces, directly by the $t$ expansion. 
%
%
%
%

Therefore, we can prescribe

\be \label{Yexp}
Y_i(\delta) = \text{Finite part} \left( \lim_{t \to +\infty} \frac{1}{t^i} \left(\delta \Li - \partial_u \theta^u (\delta) - D_a \theta^a (\delta) \right) \right) ,
\ee
for each $i$. Observe that in \eqref{Yexp} each $Y_i$ can be written as a total derivative plus a total variation. 

By taking the free function $C$ as a total derivative, $C = \partial_u X^u + D_a X^a$, we can add the last term in \eqref{Yexp} to obtain a new total derivative term. Then, the renormalized symplectic potential has the form
\be 
\theta^t_{ren}(\delta) := \theta^t(\delta) + \partial_\nu \Upsilon^{t \nu}(\delta) + \delta \Xi^t = P(\delta)(u,t,x^a)
\ee 
where $\Upsilon$ and $\Xi$ are calculated from $Y_i$, $X^u_i$ and $X^a_i$ directly, and $P$ is at most $O(t^0)$ in the $t-$expansion. This symplectic potential does not contain divergences in the $t\rightarrow \infty$ limit. The general form of the symplectic potential will be changing the upper index $t$ by a 4d index $\mu$. We have that $\Upsilon^{\mu \nu} = - \Upsilon^{\nu \mu}$, by definition of ``corner terms'' (see \cite{Freidel:2019ohg}). Without any loss of generality, we can define $\Upsilon^{jl}= 0$, for $j,l$ running in the set $\{u,x^a \}$, since these terms are not uniquely defined and do not affect the renormalization of $\theta^t$. Therefore, we have a well-defined limit

\be \label{limtheta}
\theta^\Ib_{ren}(\delta)(u,x^a) :=  \lim_{t\rightarrow +\infty} \theta^t_{ren}(\delta)(t,u,x^a) = Y_0(\delta)(u,x^a) - C (\delta)(u,x^a) 
\ee

We still have at our disposal the function $C (u,x^1,x^2)$ (the only condition we imposed so far is that it is a total derivative), which can be determined by imposing a finite limit when $u \rightarrow -\infty$ for the symplectic potential, as we will show below.

Under a general LGT, the $O(t^0)$ of the symplectic potential has $O(u^N)$ terms. Therefore, $\theta_{ren}^t$ have an expansion in powers of $u$, starting in some $u^N$ (corresponding to the highest power in $\delta$ or $\alpha$), the coefficients of the expansion depending in general on which limit we are computing, $u \rightarrow \pm \infty$. We consider the following $u$-expansion for $Y_0(\delta)$ near $u = \pm \infty$,
\be \label{Y_0}
Y_0(\delta)(u,x^a) \overset{u \rightarrow \pm \infty}{=} R_{Y_0}(\delta)(u,x^a) + \sum_{k=1}^{\infty} u^k Y^\pm_{0,k}(\delta)(x^a),
\ee 
where $\partial_u R_{Y_0}(u,x^a) = O(1/|u|^\infty)$. This condition comes from the tree-level assumption on the soft theorems and implies in particular that the limits when $u \rightarrow \pm \infty$ are in principle different,
\be
R^\pm_{Y_0}(\delta)(x^a) := \lim_{u \rightarrow \pm \infty} R_{Y_0}(\delta)(u,x^a).
\ee
By inserting \eqref{Y_0} in \eqref{limtheta}, we have
\be
\theta^\Ib_{ren} (\delta) = R_{Y_0}(\delta)(u,x^a) + \sum_{k=1}^{\infty} u^k Y^{\pm}_{0,k}(\delta)(x^a) - \partial_u X^u(\delta)(u,x^a) - D_a X^a(\delta)(u,x^a).
\ee
Observe that we can find functions $X^u,X^a$ such that their expansions around $u = \pm \infty$ renormalize the limits of the symplectic potential. For $X^u$ we find,
\be 
X^u_\pm (\delta)(u,x^a) = \sum_{k=1}^{\infty} \frac{1}{k+1} u^{k+1} Y^{\pm}_{0,k}(\delta)(x^a),
\ee
Observe that the coefficient for $u^0$ vanishes to avoid ambiguities. For $X^a$ we have,
\be \label{solving_X_a}
D_a X^a (\delta)(u,x^a) = \left\lbrace
\begin{array}{cc}
R^-_{Y_0}(\delta)(x^a) + O(1/|u|^\infty) & \text{when } u \rightarrow - \infty \\
R^+_{Y_0}(\delta)(x^a) + O(1/|u|^\infty)& \text{when } u \rightarrow + \infty 
\end{array} \right.
\ee
Finally, the symplectic potential density gives a finite result upon integration on $\Ib$ due to the fall-offs of $R_{Y_0}$.

{\bf Remark:} It is a well-known fact that the equation,
\be 
D_a V^a= 0,
\ee 
for a certain vector field $V^a$ on the sphere, has infinite solutions. In fact, it has as many solutions as there are scalar functions on the sphere (since it is a simply connected manifold). On the one hand, we have an ambiguity when solving $X^a$ in \eqref{solving_X_a}. Since $D_a X^a$ is the object that enters in the definition of $C$, such ambiguity resolves trivially in the renormalized symplectic potential, not affecting the outcome.

On the other hand, the contribution from $C$ to the symplectic potential is computed only at the boundary. Thus, it is important only for the limit value $u \rightarrow \pm \infty$ for $D_a X^a$. Then, any ambiguities in the prescription of $X_a$ are ``washed away'' by the limiting process. More on this detail in the next subsection.

As mentioned in the introduction, this renormalization is minimal because it cancels all the divergent terms while keeping \textit{unchanged} the finite ones. The linearity of the theory played a central role in the previous derivation, and also in the definition of the extended phase space. We leave for future works to delve into such a renormalization procedure in the case of non-abelian theories.

\subsection{Electric-like charge algebra}

The previous renormalization procedure adjusts exactly all the divergences while maintaining the same convergent terms discussed in subsection \ref{extended phase space and charges}. The expression for the renormalized symplectic potential is therefore:

\be
\Theta_{ren}(\delta) = \int_{\Ib^+} \theta_0(\delta) du dx^2 + \int_{S^2}  \sum_{i=1}^{\infty}  F_{ru}^{(-2-i,0)} \delta a_i dx^2 
\ee
where $\theta_0$ is the usual symplectic potential in $\F_0$. The symplectic form is the exterior derivative (in the extended phase space) of the symplectic potential:

\be \label{rnomega}
\Omega_{ren}(\delta,\delta') = \int_\Ib \omega_0(\delta,\delta')dudx^2 + \int_{S^2} \sum_{i=1}^{\infty}  \delta F_{ru}^{(-2-i,0)} \wedge \delta' a_i dx^2 
\ee
All three ingredients in the charge calculation are well defined and finite: the limit $t \to +\infty$, the integration on $\Ib$ and the series.

We can now show the full hierarchy of charges for arbitrary $O(r^n)$ LGT in QED. The electric charges associated with an LGT $\Lambda_\varepsilon$ can be calculated from \eqref{rnomega}, substituting the sequence coordinates $\{ \epsilon_i\}$. By the equation
\be
\delta Q_{\varepsilon} = \Omega_{ren}(\delta , \delta_{\Lambda_\varepsilon}),
\ee
we have
\be \label{electric_charge}
Q_\varepsilon =  \sum_{j=0}^{\infty} \int_{S^2} \sqrt{q} \epsilon_j F_{ru}^{-2-j,0} dx^2 
\ee
where we are using that $F_{\nu \mu}$ is invariant under $\delta_{\Lambda_\varepsilon}$. This expression is the same as the one obtained in \cite{Campiglia:2018dyi}.

%

Observe that the full algebra of charges is abelian:

\be 
\{ Q_{\varepsilon_1} , Q_{\varepsilon_2} \} = 0, \quad \forall \varepsilon_1 ,\varepsilon_2
\ee 
 
{\bf Remark:} The expression we found for the renormalized symplectic potential gives a symplectic form from which the electric charges $Q_{\varepsilon}$ can be obtained. Given the procedure shown in this section, some ambiguities can, in principle, spoil the conservation of the charges from $\Ib^-$ to $\Ib^+$ in a scattering process. Nevertheless, observe that the charges we obtain here are \textit{exactly} the ones given in \cite{Campiglia:2018dyi} for $\Ib^+$, and it can also be shown, via the same arguments and prescriptions, that the charges for $\Ib^-$ are,
\be \label{electric_charge_Ib_-}
Q_\varepsilon^- =  \sum_{j=0}^{\infty} \int_{S^2} \sqrt{q} \epsilon_j F_{rv}^{-2-j,0} dx^2, 
\ee
where $v$ is now the \textit{advanced} coordinate (suitable to describe $\Ib^-$), with respect to which the fall off's \eqref{falloffFru}, \eqref{decayphi}, etc., are written in analogy. Since the conservation of the charges in the classical theory was proven by Campiglia and Laddha (cf. \cite{Campiglia:2018dyi}), then any possible ambiguities in the renormalization do not affect the scattering processes between $\Ib^-$ and $\Ib^+$. It would be interesting to study if there is any impact of these ambiguities beyond the tree-level, in the non-abelian case, or in the presence of other fields.

\section{Duality extension of tower of asymptotic charges} \label{magnetictower}

In the previous sections, we treated only the electric part of Maxwell theory, renormalizing the symplectic potential in the extended phase space to contain the sub$^n$-leading charges in a natural framework. In this section, we extend the phase space (again) in order to include the magnetic freedom, \textit{á la} Freidel-Pranzetti, as in \cite{Freidel:2018fsk}. This type of extension has been thoroughly studied in recent years in several contexts: electromagnetic duality (e.g., \cite{Hosseinzadeh:2018dkh, Donnay:2020guq}), BF theories (\cite{Geiller:2021gdk}) and under more general structures (\cite{Ciambelli:2021nmv}). Throughout this section, we use differential form notation without explicitly writing the indexes to ease the notation. Also, we are considering no extra fields.

Electromagnetism possesses a \textit{duality symmetry}, which can be characterized as follows: the Lagrangian for the theory is 

\be \label{2_form_lagrangian}
\mathcal{L}[F] = \frac{1}{2} * F \wedge F,
\ee
where $\wedge$ is the wedge product in the space of $p$-forms on Minkowski space $M$ and $*$ is the Hodge dual operator, $* : \Omega^p(M) \rightarrow \Omega^{4-p}(M) $, in $M$. This operator satisfies
\be 
**\alpha = (-1)^{p(4-p) + 1}\alpha, \quad \alpha \in \Omega^p(M),
\ee
where the extra $+1$ in the exponent comes from the signature of the metric in Minkowski space. Therefore, taking $p=2$ and applying $*$ to $F$ in \eqref{2_form_lagrangian}, we have

\be 
\mathcal{L}[*F] = \frac{1}{2} *F \wedge F,
\ee
which shows the duality symmetry.

The first step in including duality symmetry is to consider the duality extension in the standard radiative phase space. On each $\Sigma_t$, we have the Freidel-Pranzetti extension for the symplectic form, \cite{Freidel:2018fsk},
\be \label{Omega_with_duality}
\Omega (\delta , \delta') = \int_{\Sigma_t} \delta A \wedge \delta' \star F + \int_{S^2} \delta a_0 \wedge \delta' B_0,
\ee
where $\star$ is the Hodge dual in the hypersurface, $a_0 \overset{S^2}{=} A + d\omega_0$ is the electric boundary gauge field, and $B_0$ is the magnetic boundary gauge field. $\omega_0$ is the \textit{edge mode}, which extends the phase space, $(A,a_0)$, which now contains this boundary field. The symplectic form now contains a corner term living in $\partial \Sigma_t$.

To make the connection with our past sections definition for $A$, we have
\be
A_{new} + d\omega = A_{old} ,
\ee 
where $old$ refers to the $A$ used in the previous sections, and $new$ is the one in the present section. In particular, the expressions for curvature tensor and the charges are still valid. Observe that $\omega$ can be thought of as a zero-order extension, using the same idea as the previous sections: extending the vector potential with a large gauge symmetry.

We distinguish between symmetries that leave fixed the bulk variable $A$ and symmetries that act only on the boundary. In the previous sections, we use this difference when defining the extension to higher order LGT, where $\delta_{\Lambda_\varepsilon}$ only acts on $A_a$ through the first component. In the present section, as it was done in \citep{Freidel:2018fsk}, we are isolating the bulk from the boundary action on the $\epsilon_0$ variation in order to have a well-defined canonical action that includes the duality symmetry, and such that the symplectic potential is invariant under the gauge transformation of the fields. 

We are working in $\Ib^+$, so in \eqref{Omega_with_duality}, we take $t \rightarrow +\infty$. The ``bulk'' part now is $A$ along $\Ib$, while the boundary is $\Ib^+_-$, with topology $S^2$. The values at the boundary are not independent since the boundary symmetries act simultaneously on both $\Ib^+_\pm$ (i.e., they are independent of $u$). Under a gauge transformation generated by $G$, both the bulk and the corner fields transform,
\be
\delta_G (A, a_0 , B_0) = ( d G, -dG , 0), 
\ee
so the variation $\delta_G$ is indeed gauge, in the sense that it has a vanishing charge $\Omega(\delta , \delta_G) = 0$, on-shell. The electric (magnetic) symmetry $\delta_{\epsilon_0} (\delta_{\lambda_0})$ acts only on the electric (magnetic) boundary field, 

\be
\delta_{\epsilon_0} (A, a_0 , B_0) = (0,d\epsilon_0 , 0),\quad \delta_{\lambda_0} (A, a_0 ,B_0) = (0, 0, d\lambda_0),
\ee
where $d\lambda_0$ is locally but not globally exact (such as in the standard examples of a charge in the $z-$axis, see section V in \cite{Freidel:2018fsk}). Observe that on-shell, upon acting with $G$, we obtain the identity 
\be
d B_0 = \star F,
\ee
which on $\Ib^+$ gives us $d B_0 = F_{ru}^{(-2,0)}$.

Our extended phase space of section \ref{extended phase space and charges} adapts well to the construction given above to the duality extension. The gauge transformation $\Lambda_{\alpha}$ is the ``bulk'' potential, generated by the boundary fields in the sequence $\alpha$, in a hierarchy graded by the correspondent power of $r$. Therefore, we can extend directly as,
\be
\Omega_{ren} (\delta , \delta') = \int_{\Ib} \left[ \delta A \wedge \delta' \star F \right]_{ren} +  \int_{S^2} \delta a_0 \wedge \delta' B_0 + \int_{S^2} \sum_{k=1}^{\infty} \delta a_k \delta' d B_k,
\ee
where $a_k$ are functions on the sphere, $B_k$ are 1-forms in the sphere locally (but not necessarily globally) exact, $a_0$ is a $1-$form \footnote{$a_0$ is not generally a gradient.}, and $ren$ indicates that is the renormalized term, given by \eqref{rnomega}. We define the action of a gauge transformations $G$ (of order $r^n$ arbitrary) as
\be
\delta_{G} A = d G , \quad \delta_G a_0  = dG_0 , \quad  \delta_{G} a_j = G_j, \quad \delta_{G} B_j =0, j \geq 1.
\ee
Evaluating the symplectic form in $\delta_G$,
\be
\Omega_{ren} (\delta , \delta_G) = - \delta \left(  \int_{\Ib} \left[ dG \wedge \star F \right]_{ren} + \int_{S^2} \delta dG_0 \wedge \delta' B_0 + \int_{S^2} \sum_{k=0}^{\infty}   G_k \delta d B_k \right),
\ee
which on-shell and after integrating by parts, we obtain (after the renormalization, allowing variations $\delta$ such that $\delta A$ has order higher than $r^0$ before taking the limit $t \rightarrow +\infty$)
\be \label{dB_value}
d B_k = F_{ru}^{(-2-k,0)}, \quad k \geq 0.
\ee
This equality establishes the value of the magnetic boundary gauge field as the field strength functions.

Finally, we will denote the magnetic variations acting on $B_k$'s as $\lambda = \{ \lambda_i \}_{i \geq 0}$, in the same fashion as we define the LGT generators. Electric (magnetic) variations act as follows on the extended phase space variables,
\be
\delta_{\epsilon_k} A =0 , \quad \delta_{\epsilon_k} a_0 = \delta_{0k} d \epsilon_k , \quad  \delta_{\epsilon_k} a_j = \delta_{kj} \epsilon_k, \quad \delta_{\epsilon_k} B_j =0, \quad k \geq 0, j\geq 1
\ee
\be
\delta_{\lambda_k} A =0 , \quad  \delta_{\lambda_k} a_j = 0,  \quad \delta_{\lambda_k} B_j =\delta_{kj} d \lambda_k , \quad k,j \geq 0,
\ee
where $d \lambda_k$ is locally but not globally defined, and $\delta_{ij}$ is Kroenecker delta.

\subsection{Charges and dual charges and their algebra}

By computing $\Omega_{ren} (\delta, \delta_{\Lambda_{\varepsilon}})$ and $\Omega_{ren} (\delta, \delta_{\Lambda_{\lambda}})$, we obtain the electric (denoted as $Q$) and magnetic (denoted as $\tilde{Q}$) charges, 

\beq
Q_{\varepsilon} &=& \sum_{k = 0}^{\infty} \int_{S^2}\epsilon_k  d B_k \\
\tilde{Q}_{\lambda} &=& \sum_{k = 0}^{\infty} \int_{S^2}  a_k d^2 \lambda_k,
\eeq
where the first integral gives directly \ref{electric_charge}, thanks to \ref{dB_value}, and the last integral does not vanish due to the failure of $d \lambda$ to be globally exact.

Finally, we have can compute the charge algebra. As the electric charges, the magnetic charges $\tilde{Q}_{\lambda}$ are abelian, 

\be
\{ \tilde{Q}_{\lambda} , \tilde{Q}_{\lambda'} \} = \delta_{\lambda } \sum_{k = 0}^{\infty} \int_{S^2} a_k d^2 \lambda'_k = 0.
\ee

The mixed Poisson bracket gives a non-trivial component of the algebra,

\be
 \{ Q_{\varepsilon} , \tilde{Q}_{\lambda} \} = \delta_{\varepsilon } \sum_{k = 0}^{\infty} \int_{S^2}a_k d^2 \lambda_k = \sum_{k = 0}^{\infty} \int_{S^2}  \epsilon_k d^2 \lambda_k =:c_k
\ee
This term shows that the boundary duality symmetry algebra possesses a hierarchy of central charges, $\{ c_k \}_{k \geq}$. We leave it to future works to analyze in detail this fact in the context of soft theorems and Ward identities.

\section{Outlook}

In this work, we obtain a well-defined symplectic form on $\Ib^+$ for the extended phase space of classical QED through a renormalization procedure from the original symplectic form, giving a derivation from first principles. With this symplectic form, the higher order LGT can be associated with the sub$^n$-leading electric charges acting canonically on the phase space. The expressions of the charges associated with the $O(r^n)$ LGT are then obtained, in agreement with the expressions previously proposed in \cite{Campiglia:2018dyi} by means of the tree-level sub$^n$-leading formulas. We compute the full electromagnetic charge algebra using the duality symmetry extension, showing a hierarchy of central extensions. Several future directions are possible in the framework of our work.

First, within the abelian theory, it would be interesting to extend the analysis to include loop corrections to the soft photon factorization formulas (\cite{Bern:2014oka, He:2014bga}). This could lead to some new structure within the charge hierarchy and between electric and magnetic charges. One of the main difficulties in this line of work is the appearance of infrared divergences. New advances in celestial CFT methods (see \cite{Pasterski:2021rjz, Guevara:2021abz, Albayrak:2020saa, Gonzalez:2020tpi, Raclariu:2021zjz}, and references therein) seem to be well suited for the incorporation of these effects. It would also be interesting, given the recent developments in the study of electromagnetic asymptotic charges at spatial infinity, such as consistently accommodating $\ln (r)$ terms (\cite{Fuentealba:2023huv}) and the study in higher dimensions (\cite{Fuentealba:2023rvf}), to establish a connection between the symplectic structure at null infinity with that at spatial infinity.  

Second, the extension to non-abelian gauge theories. In Yang-Mills theory, extending the renormalization procedure would allow us to construct a well-defined symplectic structure on an extended phase space to compute the subleading charges and their algebra. As shown in \cite{Campiglia:2021oqz}, the first step towards this is to consider a linearized extension of the phase space and restrict the charges up to $O(r)$ terms. Some progress is being made in this direction, \cite{Nagy:2022xxs}.

Finally, it would be interesting to study possible extensions of this renormalization procedure in the context of gravity. As recent works suggest, the study of higher-order diffeomorphisms seems to be a key ingredient in the extensions of the phase spaces for gravity. In \cite{Compere:2017wrj}, higher-order multipole moments generated via specific diffeomorphisms were studied and showed that they are Noether charges. In the null infinity sector it has been proposed in \cite{Campiglia:2016jdj, Campiglia:2016efb}, and worked out more recently in \cite{Freidel:2021dfs, Horn:2022acq}, that asymptotic diffeomorphisms generated by certain $O(r)$ sphere-vector fields are behind the sub-subleading soft graviton factorization \cite{Cachazo:2014fwa}. A similar idea as the one presented here could be used to identify an extended space supporting these singular transformations. 

\section*{Aknowledgements}
I would like to thank Miguel Campiglia for his comments, discussions, and feedback on the process and the final manuscript, and also to the participants of the Workshop on Celestial Symmetries, held in Montevideo in March 2022, with whom the contents of this paper were discussed, in particular Laurent Freidel, Marc Geiller, Alok Laddha, Silvia Nagy, Daniele Pranzetti and Céline Zwickel. I would also like to thank Ali Seraj and Oscar Fuentealba for their comments on the first version of the paper. I also thank the two referees for useful comments and corrections. This work was partly supported by a CAP Ph.D. fellowship, the ANII project FCE 2019-155865, and PEDECIBA.

\appendix

\section{Recursive formula for $\epsilon^{ab} F_{ab}$} \label{ap_rec}

In this appendix we prove eq. \eqref{FABrecursive}. By writing the Bianchi identities

\beq 
\partial_r F_{ab} + \partial_a F_{br} + \partial_b F_{ra} &=& 0, \\
\partial_u F_{ab} + \partial_a F_{bu} + \partial_b F_{ua} &=& 0, \\
D_c F_{ab} + D_a F_{bc} + D_b F_{ca} &=& 0,
\eeq
and taking the $u$ and $r$ derivative of the first equation, the $r$ derivative of the second one, the $D_d$ derivative of the third one and contracting with $\epsilon^{ab}$, we obtain,

\beq 
\partial_u \partial_r \epsilon^{ab} F_{ab} &=& 2 \partial_u \epsilon^{ab} D_a F_{rb} \label{App1} \\
\partial_r \partial_r \epsilon^{ab} F_{ab} &=& 2 \partial_r \epsilon^{ab} D_a F_{rb} \label{App2} \\
\partial_u \partial_r \epsilon^{ab} F_{ab} &=& 2 \partial_r \epsilon^{ab} D_a F_{ub} \label{App3} \\
D_d \epsilon^{ab} D_c F_{ab} &=& - 2 \epsilon^{ab} D_d D_a F_{bc}. \label{App4}
\eeq
In equations \eqref{App1}, \eqref{App2} and \eqref{App3} we substituted $\partial_i$ by $D_i$ for every $i=a,b$, since we are contracting with $\epsilon^{ab}$. 

Using identities

\be 
D_a D_b F_{cd} = D_b D_a F_{cd} - q^{ef}R_{ecab} F_{fd} - q^{ef} R_{edab} F_{cf}, \quad R_{abcd} = \frac{R}{2} (q_{ac}q_{bd} - q_{ad}q_{bc}),
\ee
we have
\be 
D^d D_a F_{bd} = D_a D^d F_{bd} + R F_{ab}.
\ee

By contracting \eqref{App4} with $q^{cd}$ and u, and the previous equation,

\be 
\Delta \epsilon^{ab} F_{ab} = - 2 \epsilon^{ab} D_a D^d F_{bd} - 2 R \epsilon^{ab} F_{ab}.
\ee

Next, consider Maxwell equation \eqref{Maxa}, and take de $D_d$ derivative and contract with $\epsilon^{da}$,

\be 
\epsilon^{da} D_d j_a = -\partial_r \epsilon^{da} D_d (F_{ua} - F_{ra})+ \partial_u  \epsilon^{da} D_d F_{ra} + \frac{1}{r^2} \epsilon^{da} D_d D^b F_{ab}.
\ee

Substituting the previous equations, we arrive at 

\be 
2 \epsilon^{ab} D_a j_b = 2\partial_u \partial_r \epsilon^{ab} F_{ab} - \partial_r \partial_r \epsilon^{ab} F_{ab} - \frac{1}{r^2} (\Delta \epsilon^{ab} F_{ab} + 2 R \epsilon^{ab} F_{ab}).
\ee
where $R =2$, is the scalar curvature of $q_{ab}$.

\providecommand{\href}[2]{#2}\begingroup\raggedright\endgroup

\end{document}